\documentclass[a4paper,fleqn]{cas-sc}

\usepackage[numbers]{natbib} 

\usepackage{eurosym}
\usepackage{graphicx}
\usepackage{siunitx}

\def\tsc#1{\csdef{#1}{\textsc{\lowercase{#1}}\xspace}}
\tsc{WGM}
\tsc{QE}
\tsc{EP}
\tsc{PMS}
\tsc{BEC}
\tsc{DE}

\shorttitle{Fused silica wheel trap by SLE}
\shortauthors{J. Vejrosta et al.}
\begin{document}

\title[mode = title]{Microfabrication of Paul wheel trap chip in fused silica by selective laser etching}  




%

\author[1,2]{Vejrosta}[type=editor,
                        auid=000,bioid=1,
                        prefix=Jakub]
\cormark[1]
\fnmark[1]
\ead{jakub.vejrosta@isibrno.cz}
\affiliation[1]{organization={Institute of
Scientific Instruments of the Czech Academy of Sciences},
    addressline={Královopolská 147}, 
    city={Brno},
    postcode={612 00}, 
    country={Czech Republic}}
\author[1,2]{Šilhan}[prefix=Lukáš]
\fnmark[1]

\author[1]{Šerý}[prefix=Mojmír]

\affiliation[2]{organization={Institute of Physical Engineering, Faculty of Mechanical Engineering, Brno University of Technology},
    addressline={Technická 2896/2}, 
    city={Brno},
    postcode={616 69}, 
    country={Czech Republic}}
\author[1]{Fořt}[prefix=Tomáš]

\author[1]{Brzobohatý}[prefix=Oto]

\author[1]{Zemánek}[prefix=Pavel]

\cortext[cor1]{Corresponding author}
\fntext[1]{These authors contributed equally to this work.}

\begin{abstract}
We present a rapid, mask-free fabrication workflow for the Paul wheel trap chip based on selective laser etching (SLE) of fused silica. The complete process - femtosecond laser writing, wet-chemical etching in KOH ($\sim$230~$\mu$m/h, releasing the trap geometry from a 500~$\mu$m wafer in approximately 2~h), RF magnetron sputtering of silver electrodes, and direct femtosecond laser ablation for electrode patterning - is carried out in-house, bypassing photolithographic processing and custom shadow masks entirely. Nine chips per 4-inch wafer can be produced in a single working day, reducing the fabrication lead time by more than one order of magnitude relative to the outsourced CVD diamond design. The functionality of the fabricated wheel trap chip is confirmed by a proof-of-concept microparticle trapping demonstration, in which a single microparticle was confined continuously for 72~h at atmospheric pressure. This establishes SLE of fused silica as a rapid, economically accessible fabrication platform for Paul traps in levitated-optomechanics applications.
\end{abstract}




\maketitle

\section{Introduction}

The rapid progress in quantum science and technology has driven a growing demand for precisely engineered hardware capable of confining and manipulating individual charged particles \cite{PhysRevLett.62.403}. Beyond atomic-ion quantum computing \cite{PRXQuantum.3.010347}, radio-frequency (RF) Paul traps have found increasing application in levitated optomechanics, where charged micro- and nanoparticles serve as ultrasensitive force sensors and as platforms for probing quantum mechanics at mesoscopic scales \cite{bonvin2024state}.

The so-called wheel trap chip, an open-geometry planar Paul trap developed at NIST, provides excellent optical access and well-defined trapping fields \cite{Chen2017thesis}. The original design was fabricated from chemical vapor deposited (CVD) diamond by laser cutting and metallized with sputtered gold electrodes \cite{Brewer2019}. Diamond was selected primarily for its high thermal conductivity, which mitigates resistive heating under the strong RF drive used in atomic-ion experiments. However, in levitated optomechanics the RF drive frequencies and powers are substantially lower, and the resulting thermal load on the trap is correspondingly reduced \cite{Millen_2020}, making the high thermal conductivity of diamond unnecessary and lower-cost dielectric substrates such as fused silica a viable alternative. Moreover, the established diamond process carries substantial costs and lead times - outsourced fabrication of laser-cut CVD diamond wheel trap chips with sputtered gold metallization typically requires several weeks, and the laser cutting itself introduces heat-affected zones, limited feature precision, and potential carbon residue contamination between electrodes \cite{Setzer2022}. Established microfabrication techniques such as CNC machining \cite{KRIMPENIS2022328}, photolithography \cite{Baracu_2021}, and direct laser ablation \cite{Piqué_2016} each impose their own constraints in terms of achievable tolerances, surface quality, processing complexity, or geometric flexibility \cite{mi15101274,verma2026recent}. More recently, the wheel trap chip geometry has been realized in fused silica for levitated optomechanics, where the substantially lower RF drive compared to atomic-ion experiments makes the lower thermal conductivity of fused silica acceptable \cite{Bonvin2024}. In that work, the trap was micro-machined and gold-coated by a commercial vendor.

Selective laser etching (SLE), also referred to as femtosecond laser-induced selective etching (FLISE), provides a compelling alternative \cite{Butkute:21}. In this two-step subtractive process, focused femtosecond laser pulses locally modify the bulk of a transparent dielectric, and subsequent wet-chemical etching selectively removes the modified regions \cite{articleUPBMT}. SLE enables truly three-dimensional geometries with micrometer-scale precision, vertical sidewalls, aspect ratios exceeding $100{:}1$, and near-optical surface quality without heat-affected zones \cite{Butkute2022}. Among the materials amenable to SLE, fused silica ($a$-SiO$_2$) is the most established and well-characterized substrate \cite{DEMICHELE2022121384}. Its amorphous nature ensures isotropic laser-matter interaction and etching behavior, eliminating the crystal-orientation dependencies that complicate processing of crystalline substrates \cite{10.1063/5.0228383}. Fused silica further offers wide optical transparency from the UV to the near-IR, low dielectric loss, and high chemical inertness, as well as high vacuum compatibility\cite{JAQUET2021100060}. For the kHz-range RF drives and modest voltages typical of levitated-optomechanics traps, the resistive and dielectric heating of the trap structure is negligible (Section~\ref{sec:comparison}), so the low thermal conductivity of fused silica relative to CVD diamond imposes no practical limitation. This has recently been confirmed experimentally: a fused silica wheel trap chip was successfully operated for levitated optomechanics by Bonvin et al. \cite{Bonvin2024}, establishing fused silica as a viable substrate for this trap geometry. Being available as high-quality optical-grade wafers at a small fraction of the cost of CVD diamond, and being the most established substrate for selective laser etching \cite{Marcinkevicius:01}, fused silica is a natural choice for a low-cost, rapid wheel-trap fabrication route. The substrate choice is also dictated by the process itself. SLE relies on bulk modification by focused pulses at 1030~nm, to which fused silica is transparent \cite{Marcinkevicius:01}. In contrast, silicon absorbs linearly at this wavelength ($\alpha \approx 30$~cm$^{-1}$ at 1.03~$\mu$m \cite{green2008}), so that the modification cannot be confined to the focal volume in the bulk; laser-induced selective etching of silicon has been demonstrated with sources near 1.5~$\mu$m \cite{tokel2017}, but not at the wafer thickness, selectivity and process maturity available for fused silica. Fused silica additionally provides the electrical insulation required to carry separated electrodes without an interlayer \cite{dietl2025}.

Building on this, we present a fabrication-oriented route for the NIST-type wheel trap chip \cite{Chen2017thesis} based on SLE of fused silica, aimed at maximizing the simplicity of fabrication while minimizing fabrication lead time and complexity. Both the trap geometry and the electrode patterning are produced on a single femtosecond laser workstation, with the electrodes defined by direct laser ablation of a sputtered silver film. The process requires no photolithography, no custom shadow masks, and no specialized electrode-shadowing geometries, reducing the wheel trap chip to a short sequence of standard microfabrication steps. The advantage over a photolithographic route lies not in the speed of any individual step, where photolithography is competitive, but in the number of steps and the absence of any external fabrication dependency: no mask design, mask procurement or lithographic patterning is required at any stage. The argument is also geometric: the electrodes facing the trapping region are the vertical sidewalls of released blades, which spin-coated resist and top-down exposure do not pattern, whereas here they are metallized by blanket sputtering from both wafer faces and segmented by laser ablation with the chip tilted. This simplicity allows nine chips to be produced from a single 4-inch wafer in one working day using equipment common to most microfabrication facilities. A proof-of-concept microparticle trapping experiment confirms device functionality; detailed characterization of trap dynamics is beyond the scope of this fabrication-oriented work and will be reported separately.
Related fabrication routes for dielectric traps have been reported: stacked fused silica traps produced by laser-enhanced etching with shadow-mask metallization \cite{ragg2019}, monolithic structures relying on self-shadowing trenches for electrode isolation \cite{araneda2020,teh2024}, and post-metallization laser ablation for electrode patterning \cite{decaroli2021}; more recent selective-laser-etched traps combine laser-defined isolation structures with evaporative or sputtered metallization \cite{brucke2026,schroer2026}. We therefore claim neither selective laser etching, nor mask-free electrode isolation, nor post-metallization laser ablation as individually new techniques. The contribution of this work is their combination into a single-workstation route for the wheel trap geometry, in which the structural geometry and the electrode pattern are defined in two separate direct-write steps, so that the electrical layout is set on the already-released structure rather than fixed during deposition. 

\section{Experimental techniques and materials}

\subsection{Materials}

Fused silica wafers (double-side polished, thickness $500~\mu$m, JGS2 grade) were obtained from Microchemicals GmbH. For wet-chemical etching, an aqueous potassium hydroxide solution was prepared by dissolving KOH pellets (Penta Chemicals) in deionized water at a concentration of $7.5$~M (33~wt\%).

\subsection{Micromachining workstation}

\begin{figure}[pos=!htbp]
	\centering
	\includegraphics[width=0.8\textwidth]{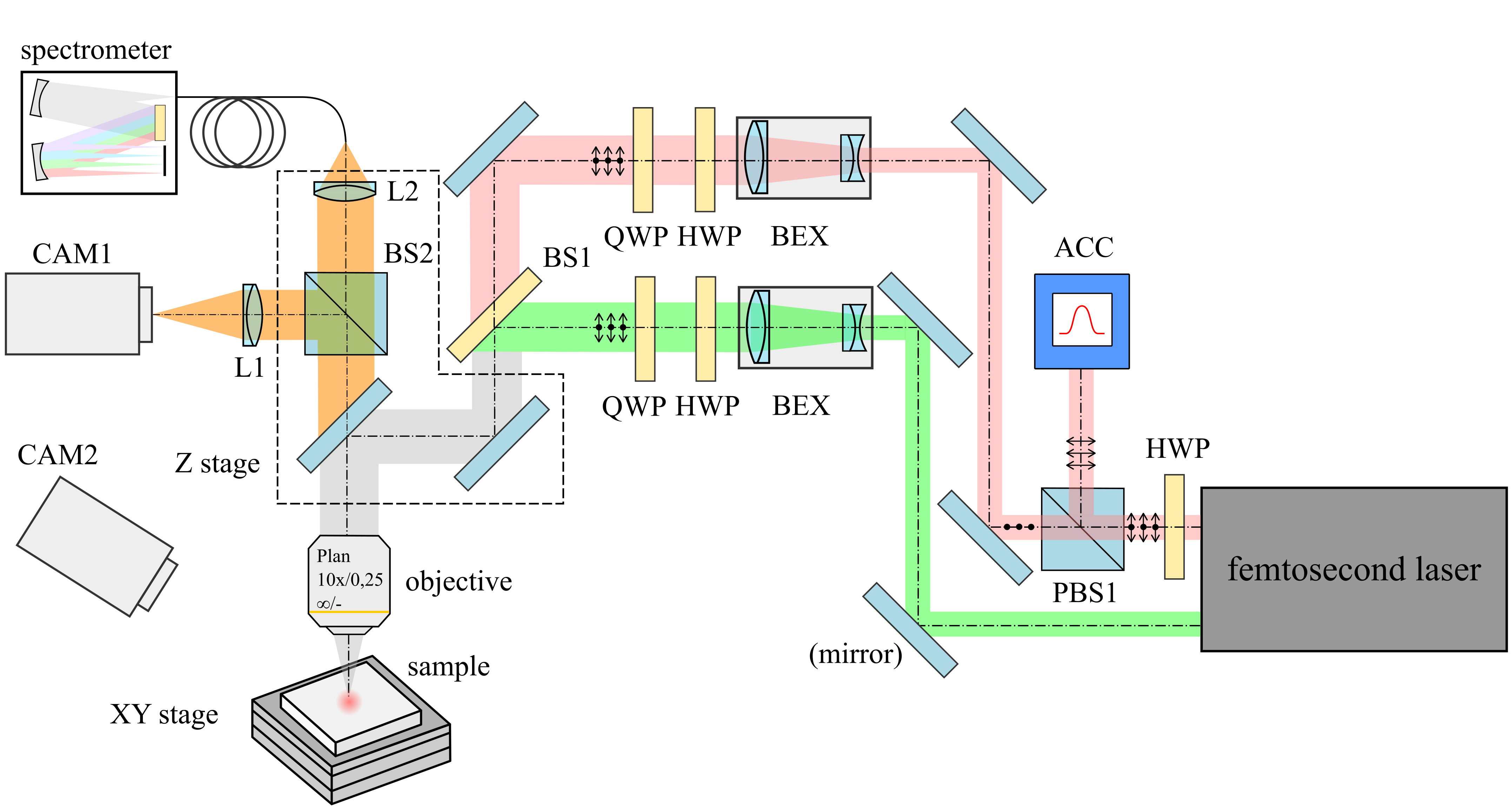}
	\caption{Optical layout of the custom-built micromachining station. PBS - polarizing beam splitter, HWP - half-wave plate, QWP - quarter-wave plate, BEX - beam expander, BS - beam splitter, L - lens, CAM - camera. The focusing objective is interchangeable depending on the application.}
	\label{fig:setup}
\end{figure}

Femtosecond laser structuring was performed on a custom-built micromachining station. The optical layout is shown in Fig.~\ref{fig:setup}. The laser source is a Jasper X0 (Fluence Technology, Poland), delivering pulses at the fundamental wavelength of 1030~nm with a pulse energy up to 200 $\mu$J, repetition rate up to 2~MHz, and pulse duration adjustable from 330~fs to 20~ps. 
The laser beam is guided through a periscope assembly using dielectric mirrors (Eksma Optics) and polarization control optics (half-wave and quarter-wave retarders) to the focusing objective.
The setup supports both the fundamental (1030~nm) and second-harmonic (515~nm) wavelengths, combined via a dichroic mirror. For selective laser etching, a $20\times$ microscope objective (Olympus, NA~=~0.40) was used to focus the beam into the substrate volume. To exploit the full numerical aperture of the objective and achieve the smallest possible focal spot, the beam was expanded by the beam expander (BEX) to fill the back aperture of the objective. For direct laser ablation of thin films, a plano-convex lens with a focal length of 100~mm was used instead, providing a larger spot size. Two cameras are used for alignment: one images the back-reflected beam to locate the focus on the top and bottom surfaces of the wafer, defining the wafer plane in $z$, while the other provides a top-view image of the wafer for positioning on the $XY$ stage. An autocorrelator (ACC) and a spectrometer are integrated into the beam path for monitoring the pulse duration and spectrum, respectively.

The sample is mounted on a ceramic vacuum chuck attached to a precision XYZ gantry stage built on a granite base for vibration isolation. Individual laser pulses are triggered via TTL signals from a laser control interface (LCI) module synchronized with the stage position, enabling precise control over the spatial distribution of laser-modified zones within the substrate.

\subsection{Sample preparation}
\label{sec:sample_prep}

\subsubsection{Selective laser etching}
\label{sec:sle_prep}
Substrates were laser-structured by scanning with focused femtosecond beam through the wafer volume according to the wheel trap chip geometry. Focused femtosecond pulses deposit energy in the focal volume through nonlinear
(multiphoton and avalanche) ionization, locally modifying the fused silica by breaking Si--O bonds and forming self-organized nanogratings. This modification both loosens the glass network and creates nanoporous channels that promote etchant infiltration, rendering the modified material far more susceptible to wet-chemical attack than pristine fused silica~\cite{Qi:18, Hermans2014SelectiveLE}. Rather than modifying the full volume to be removed, only the contours of the trap geometry were written: the outline of each blade electrode and the surrounding clearance was traced as a stack of contours spaced 5~$\mu$m apart in $z$, spanning the entire 500~$\mu$m wafer thickness. Subsequent etching attacks these through-thickness modified walls, releasing the enclosed central regions as intact pieces once the contours are fully undercut, without the need to modify the interior material.

The laser writing parameters - pulse energy, repetition rate, scanning speed, and polarization state - were optimized through systematic writing tests (see Section~\ref{sec:sle_results}). Circular polarization was used throughout to suppress the formation of polarization-dependent anisotropic nanogratings, which otherwise make the etch rate sensitive to the angle between the writing polarization and the scanning direction; this ensures a uniform etch rate and selectivity for arbitrarily oriented features in the wafer plane~\cite{LI2019188, mi11050483}.

\begin{figure}[pos=!htbp]
  \centering
  \includegraphics[width=0.5\linewidth]{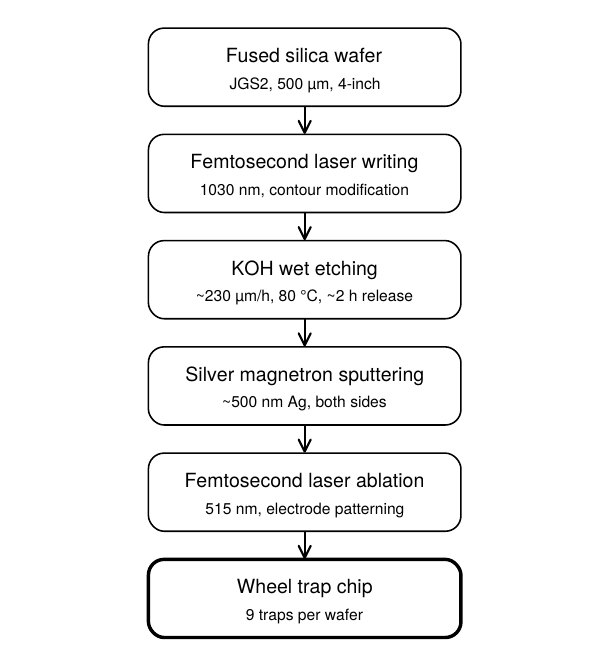}
  \caption{Fabrication workflow of the fused silica wheel trap chip, from the JGS2 wafer to the finished chip: femtosecond laser writing of the
  trap contours, KOH wet etching to release the geometry, silver
  magnetron sputtering, and femtosecond laser ablation for electrode
  patterning.}
  \label{fig:workflow}
\end{figure}

Laser-modified fused silica substrates were etched in ${\sim}7.5$~M KOH solution at $80~^\circ$C with point-source ultrasonic agitation. The ultrasonic agitation enhances etchant transport into the high-aspect-ratio laser-modified channels, improving etching uniformity and rate. The process exhibited high selectivity, with minimal attack on unmodified regions, and released the wheel trap chip geometry from the 500~$\mu$m thick wafer within approximately \SI{2}{\hour}. KOH was chosen over HF for its substantially higher etch selectivity and its lower hazard in routine use~\cite{Hermans2014SelectiveLE}. The complete fabrication sequence is summarized in Fig.~\ref{fig:workflow}.

\subsubsection{Thin film deposition and electrode patterning}
\label{sec:ablatin}
Prior to metallization, the fused silica substrates were cleaned by a
standard RCA procedure~\cite{Kern_1990}, comprising the SC-1
(NH\textsubscript{4}OH:H\textsubscript{2}O\textsubscript{2}:H\textsubscript{2}O)
and SC-2 (HCl:H\textsubscript{2}O\textsubscript{2}:H\textsubscript{2}O) steps to
remove organic and ionic contaminants, respectively. Substrates were rinsed in
flowing deionized water between steps and dried under a nitrogen stream.

\begin{figure}[pos=!htbp]
	\centering
	\includegraphics[width=16cm]{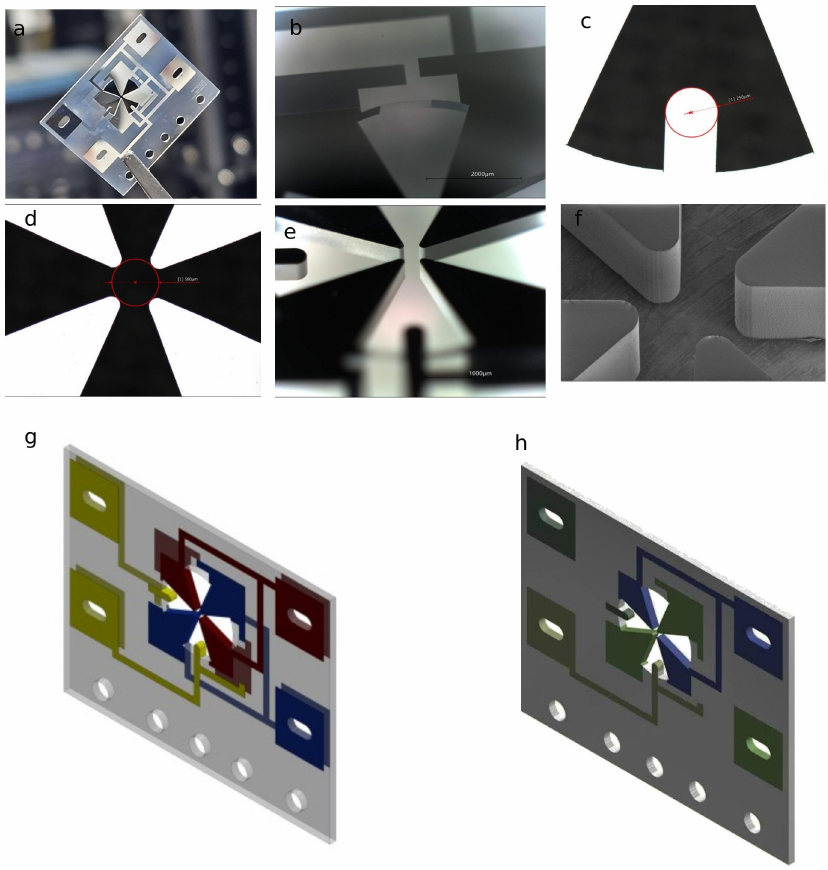}
	\caption{Overview of the fabricated fused silica wheel trap chip and its components. 
    (a) Photograph of the manufactured chip with silver electrode metallization. 
    (b) Optical micrograph of the laser-ablated electrode pattern; dark regions are 
    ablated areas where the silver film was removed, defining the isolation gaps between 
    electrodes. 
    (c) Dimensional tolerancing of the compensation electrode. 
    (d) Dimensional tolerancing of the trapping region between the four blade electrodes. 
    (e) Optical micrograph along the trap axis showing the four blade electrodes 
    converging toward the trapping region. (f) SEM image of the trapping region.
    (g) 3D rendering of the present fused silica wheel trap chip design.
    (h) 3D rendering of the original NIST wheel trap design for comparison.} 
	\label{fig:all}
\end{figure}

Silver thin films were deposited by RF magnetron sputtering from a silver target (silver 99.99\%, SAFINA) at 500~W RF power in an argon atmosphere (20~sccm, working pressure 0.987~Pa) with a base pressure of $6.6 \times 10^{-5}$~Pa. No dedicated adhesion layer was used; instead, substrates were cleaned by RF argon ion etching (150~W, 600~s, 0.140~Pa, $U_\mathrm{sub} = -70$~V) immediately prior to deposition to promote film adhesion on the fused silica surface. Reference samples prepared by the same system and process without an adhesion layer show no visible tarnishing after ten years of ambient storage, consistent with reported two-year monitoring in which no silver compounds were detected \cite{baburin2020}; omitting a Ti or Cr layer is advantageous here, since it would otherwise be exposed at the ablated isolation gaps. The silver film (${\sim}500$~nm on the wafer faces, determined from the deposition rate and deposition time) was deposited on both sides of the substrate in a stationary configuration without a homogenization shield. The comparatively non-directional nature of magnetron sputtering also provides metallization of the blade-electrodes sidewalls, which SEM and confocal imaging confirm to be continuously coated, ensuring electrical continuity of the electrodes around the full blade profile (Fig.~\ref{fig:all}c,d). A tilt-angle calibration of the sputtering system gives 43\% of the nominal thickness at 80$^{\circ}$ incidence and an estimated 23\% at 90$^{\circ}$, so that two depositions of 500~nm nominal thickness, one from each wafer face, leave approximately 200~nm on the sidewall. This figure is an estimate, as the calibration was performed with a different target material and no cross-section was measured; continuity was verified directly by a two-point resistance of 1.5~$\Omega$ measured through the sidewall from one wafer face to the other.

Electrode patterns and electrical wiring were defined by direct femtosecond laser ablation of the deposited thin films on the same micromachining workstation. Because laser ablation involves a large number of coupled parameters (pulse energy, repetition rate, scanning velocity, hatch spacing,
focusing condition, and wavelength), several were fixed from preliminary tests to reduce the optimization space. The second harmonic at 515~nm was used for its higher absorption in the metallic film relative to the fundamental at 1030~nm, and focusing was performed with a plano-convex lens ($f = 100$~mm) at a defocus of $-1.2$~mm, giving a larger spot size than the SLE objective and hence a wider single-pass ablation track. The remaining parameters were then optimized for clean removal of the silver film without damaging the underlying fused silica or the metal film on the opposite face. The final ablation used a pulse energy of 300~nJ, a hatch spacing of 7.5~$\mu$m, a pulse-to-pulse spacing of 500~nm, and a cross-hatch double-pass strategy at a scanning velocity of 100~mm/s. The isolation gaps on the blade sidewalls were patterned in the same step with the chip tilted by 60$^{\circ}$ about the trap axis relative to the beam, so that the focused beam reached the sidewall surface at otherwise identical ablation parameters; the four sidewall segments were patterned sequentially by rotating the chip between exposures. The complete fabricated device, together with characterization  of the electrode sidewalls, electrode geometry tolerancing, and the trapping region, is summarized in Fig.~\ref{fig:all}.

\subsection{Microparticle trapping}
\label{sec:trap_setup}
\begin{figure}[pos=!htbp]
	\centering
	\includegraphics[width=16cm]{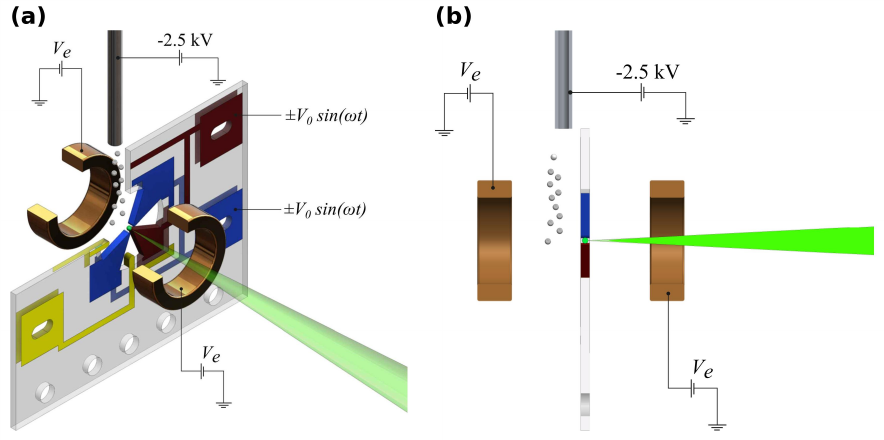}
    \caption{Operation of the wheel trap chip. (a) Schematic of the trap in operation. The four converging blade electrodes form the RF quadrupole: opposing pairs (blue and red) are symmetrically driven at $\pm V_0\sin(\omega t)$, generating the radial confining pseudopotential. The outer electrodes (yellow) are held at a DC compensation voltage $V_e$. The upper needle electrode (biased at $-2.5$~kV) loads charged microparticles into the trapping region by electrospray. The trapped microparticle (green) at the trap center is illuminated by a 532~nm laser beam. (b) Cross-section of the assembled device along the trap axis (chip metallization not resolved at this scale), showing the fused-silica chip, the copper tube endcap electrodes at $V_e$ providing axial confinement, the electrospray needle at $-2.5$~kV, the electrospray plume, and the 532~nm illumination beam.}
    \label{fig:trappingsetup}
\end{figure}

The operating principle of the wheel trap is illustrated in Fig.~\ref{fig:trappingsetup}: the four converging blade electrodes form the RF quadrupole, with opposing pairs driven in antiphase at $\pm V_0\sin(\omega t)$ to generate the radial confining pseudopotential, while the outer electrodes are held at a DC compensation voltage $V_e$ to cancel stray fields and minimize micromotion. Axial confinement is provided by two copper endcap electrodes, and charged microparticles are delivered into the trapping region by an electrospray needle.
The fabricated wheel trap chip was mounted inside a 3D-printed enclosure equipped with electrical feedthroughs. Two endcap electrodes fabricated from copper tubes (inner diameter 5~mm, wall thickness 1~mm) were integrated into the assembly to provide axial confinement. Microparticles were loaded by electrospraying a solution of 0.95~$\mu$m polystyrene microspheres dispersed in isopropyl alcohol; the acceleration and charging voltage applied to the nozzle was approximately 2.5~kV relative to ground. A sinusoidal driving signal in the range of 10--20~kHz was generated using a data acquisition card (NI~USB-6351) and amplified by an audio amplifier (Stax~SRM-353X), reaching peak-to-peak voltages up to 490~V on the trap blades.

\section{Results and Discussion}

\subsection{Selective laser etching}
\label{sec:sle_results}
\begin{figure}[pos=!htbp]
	\centering
	\includegraphics[width=8cm]{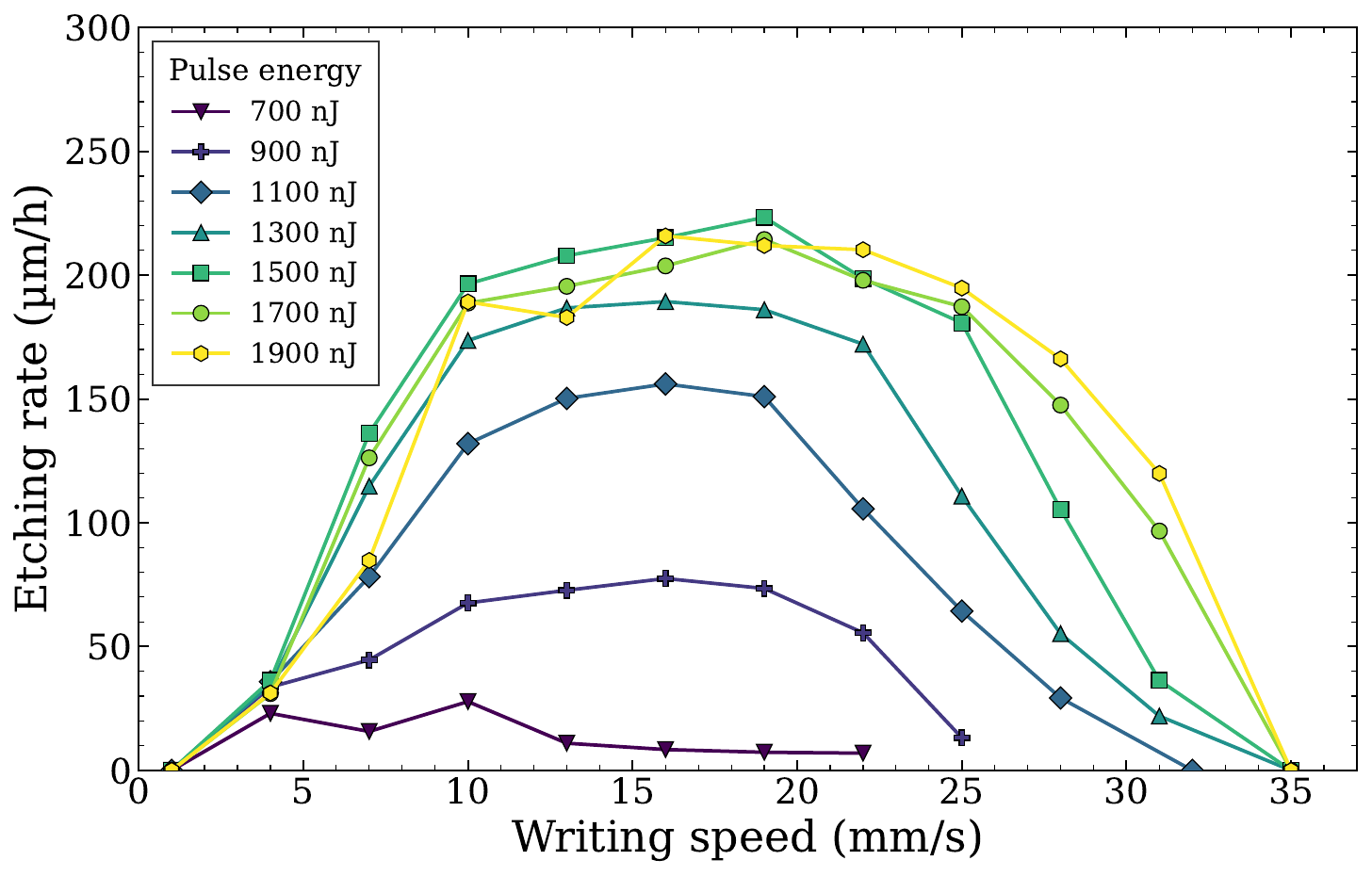}
	\caption{Dependence of the etch rate on pulse energy and scanning velocity for circularly polarized pulses at 50~kHz repetition rate.}
	\label{fig:etch_rate}
\end{figure}

The dependence of the etch rate on the laser writing parameters is summarized
in Fig.~\ref{fig:etch_rate}. The etch rate increases with pulse energy and
exhibits a pronounced maximum as a function of writing speed, defining a
parameter window suitable for releasing the wheel trap chip geometry. Under the
selected conditions (\SI{1500}{\nano\joule} pulse energy,
\SI{20}{\milli\meter\per\second} writing speed, \SI{50}{\kilo\hertz}
repetition rate), an etch rate of 230~$\mu$m/h was obtained, sufficient
to release the trap geometry from the 500~$\mu$m wafer within
${\sim}2$~h of wet-chemical etching in KOH. SEM characterization of the etched
structure (Fig.~\ref{fig:all}f) confirms vertical sidewalls and well-defined
geometry. The obtained Sa microroughness after the wet etching was \SI{0.69}{\micro\meter} on the vertical plane.

This roughness and the $\pm5~\mu$m form tolerance specified for the blade geometry act on different scales. The $\pm5~\mu$m form tolerance is referenced to a trapping aperture of 500~$\mu$m diameter (Fig.~\ref{fig:all}d), i.e. $r_{0} \approx 250~\mu$m, and therefore corresponds to approximately 2\% of the electrode-to-trap distance; deviations on this scale displace the RF null and add higher-order field components \cite{berkeland1998}, and this is the operative geometric specification of the trap. The measured $S_{a}$ is smaller by a further factor of seven, amounting to approximately 0.3\% of $r_{0}$, and is a short-wavelength surface texture rather than a deviation of the blade form; since a boundary perturbation decays with distance from the electrode at a rate set by its lateral wavelength \cite{jackson1999}, its contribution at the trapping site is bounded above by this amplitude ratio and is therefore negligible relative to the form tolerance. For comparison, laser-etched fused silica traps with a sidewall roughness below 100~nm and an ion-electrode distance of ${\sim}100~\mu$m, i.e. a ratio of order 0.1\%, have been operated successfully \cite{araneda2020}; the ratio of 0.3\% in the present device is of the same order, and the demonstrated confinement of a microparticle for 72~h is consistent with this. No numerical field error is assigned here, as that would require an electrostatic model of the measured geometry.

\subsection{Direct laser ablation of thin films}
\label{sec:ablation_results}
The optimized ablation parameters (Section~\ref{sec:ablatin}) produced
clean removal of the silver film with well-defined isolation gaps between
adjacent electrodes, without visible damage to the underlying fused silica or
to the metallization on the opposite face. The resulting electrode pattern is
shown in Fig.~\ref{fig:all}b.

Nine chips were processed on a single 4-inch wafer. One was lost to handling damage during etching, leaving eight for metallization; no further chips were lost during sputtering or laser patterning. Five of the eight have been electrode-patterned, all showing continuous metallization and clean isolation gaps under optical inspection, and the remaining three are held in reserve for process development. Two of the five were assembled and tested, and both trapped microparticles. We report these stage-specific counts rather than an overall functional yield, since three chips remain unpatterned and three patterned devices untested.

\subsection{Comparison with existing designs}
\label{sec:comparison}
We compare the fused silica SLE route presented here with the original NIST CVD diamond wheel trap chip \cite{Chen2017thesis} and the commercially fabricated fused silica chip of Bonvin et al.\ \cite{Bonvin2024}. The trap geometry presented here was fabricated in 500~$\mu$m JGS2 fused silica to a specified tolerance of $\pm$5~$\mu$m, with silver blade electrodes ($\sim$500~nm) defined by direct femtosecond laser ablation. The principal advantage of the proposed route is the simplicity and speed of the process: nine chips are produced from a single 4-inch wafer in one working day on a single femtosecond laser workstation, using a short sequence of standard microfabrication steps and no custom masks, whereas the CVD diamond design yields single devices with fabrication lead times of several weeks. The present chip design and the original NIST geometry are shown side by side in Fig.~\ref{fig:all}(g) and (h), respectively. The principal trade-off is the substantially lower thermal conductivity of fused silica compared to CVD diamond. For a capacitive load the ohmic and dielectric losses are $P_{\mathrm{ohm}} = (2\pi f C V_{\mathrm{rms}})^{2}R$ and $P_{\mathrm{diel}} = 2\pi f C V_{\mathrm{rms}}^{2}\tan\delta$ \cite{dietl2025}, where $f$ is the RF drive frequency, $C$ the electrode capacitance, $V_{\mathrm{rms}}$ the root-mean-square drive voltage, $R$ the electrode resistance and $\tan\delta$ the dielectric loss tangent of the substrate. With $f = 20$~kHz, $V_{\mathrm{rms}} = 173$~V (Section~\ref{sec:trap_setup}), $C \approx 1$~pF, the measured $R = 1.5~\Omega$ and $\tan\delta = 5\times10^{-4}$ \cite{heraeus}, this gives $P_{\mathrm{diel}} \approx 1.9~\mu$W and $P_{\mathrm{ohm}} \approx 0.7$~nW. Raising $C$ and $\tan\delta$ each by a factor of ten and taking the drive amplitude between antiphase electrodes keeps the total below a milliwatt, well under the tens of milliwatts typical of MHz-driven atomic-ion traps \cite{dietl2025}. The low thermal conductivity of fused silica therefore imposes no practical limitation at this drive. The electrode thickness is likewise not a limiting parameter at this drive: the silver skin depth at 20~kHz is approximately 450~$\mu$m \cite{jackson1999}, far exceeding the deposited film, so the ${\sim}200$~nm sidewall metallization carries current uniformly across its full cross-section, and the two-point resistance of 1.5~$\Omega$ measured through the sidewall (Section~\ref{sec:ablatin}) confirms continuity around the blade profile.

Whereas the fused silica wheel trap chip of Bonvin et al.\ \cite{Bonvin2024} established fused silica as a viable substrate for this geometry but was produced by an outsourced commercial process, the present work realizes both the trap geometry and the electrode pattern in-house on a single femtosecond laser workstation, requiring no photolithography and no custom masks.

\begin{figure}[pos=!htbp]
	\centering
	\includegraphics[width=7cm]{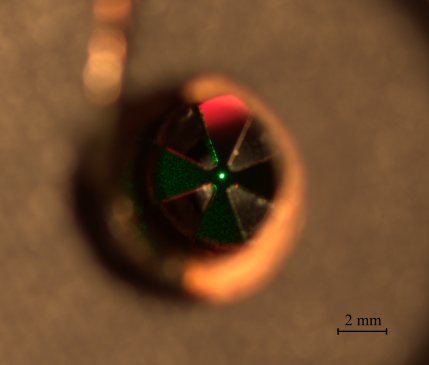}
	\caption{Trapped 0.95 $\mu$m polystyrene microparticle (green spot) confined at the center of the wheel trap chip, illuminated by 532 nm laser.}
	\label{fig:trapping}
\end{figure}

\subsection{Microparticle trapping demonstration}
\label{sec:trap}
The functionality of the fabricated wheel trap chip was verified through microparticle trapping experiments, following the operating principle illustrated in Fig.~\ref{fig:trappingsetup}. Trapped microparticles were identified by 532~nm laser light scattered from the trapping region, shown in Fig.~\ref{fig:trapping}: the microparticle is confined in the central region between the four blade electrodes, where it scatters the incident light, confirming successful operation of the device at atmospheric pressure. The geometry of the fabricated trap, viewed along the trap axis, is shown in Fig.~\ref{fig:all}e, and a photograph of the completed device in Fig.~\ref{fig:all}a. Trapping stability was assessed by holding a single 0.95~$\mu$m polystyrene microparticle continuously for 72~h at atmospheric pressure under unchanged drive conditions, with no loss of confinement and no visible degradation of the blades or silver electrodes on subsequent inspection. The experiment was terminated deliberately, so this figure is a lower bound rather than a measured limit.

\section{Conclusion}
We have demonstrated a fabrication route for the Paul wheel trap chip based on selective laser etching of fused silica, with substantially reduced fabrication time and process complexity relative to the original CVD diamond design. The femtosecond laser writing, KOH wet-chemical etching, and direct laser ablation for electrode patterning are carried out as a short sequence of standard microfabrication steps, with the only additional step being RF magnetron sputtering of the silver electrodes, without photolithographic processing or custom masks, yielding nine chips per 4-inch wafer in a single working day. A proof-of-concept microparticle trapping demonstration confirms basic device functionality; detailed characterization of trap dynamics will be reported in a forthcoming study. This work establishes SLE of fused silica as a simple and rapid fabrication platform for MEMS-based particle traps in quantum and levitated-optomechanics experiments.


\section*{Data availability}

All research data reported in this study are available upon request.

\section*{Conflict of interest}

The authors declare that they have no known competing financial interests or personal relationships that could have appeared to influence the work reported in this paper.

\printcredits

\section*{Acknowledgments}

This work was supported by the project Quantum Engineering and Nanotechnology (CZ.02.01.01/00/22\_008/0004649). The authors thank T.~Motlov\'{a} for scanning electron microscopy characterization of micromachined parts. The authors also acknowledge the institutional support of the Czech Academy of Sciences (RVO:~68081731).

\section*{CRediT authorship contribution statement}

\textbf{J.~Vejrosta:} Conceptualization, Methodology, Investigation, Software, Visualization, Writing - original draft.
\textbf{L.~\v{S}ilhan:} Conceptualization, Methodology, Investigation, Writing - original draft.
\textbf{M.~\v{S}er\'{y}:} Conceptualization, Supervision, Funding acquisition, Writing - review \& editing.
\textbf{T.~Fo\v{r}t:} Investigation (thin film deposition).
\textbf{O.~Brzobohat\'{y}:} Investigation, Writing - review \& editing.
\textbf{P.~Zem\'{a}nek:} Supervision, Funding acquisition.

\end{document}